\def\a{\alpha}
\def\f{\phi}\def\h{\theta}
\def\l{\lambda}\def\m{\mu}\def\n{\nu}\def\r{\rho}

\def\O{\Omega}

\def\na{\nabla}
\def\mo{{-1}}\def\ha{{1\over 2}}

\def\mn{{\mu\nu}}

\def\af{asymptotically flat }
\def\fe{field equations }\def\bh{black hole }\def\as{asymptotically }

\def\bhs{black holes }

\def\cco{cosmological constant }
\def\ssy{spherically symmetric }

\def\ads{anti-de Sitter }
\def\RN{Reissner-Nordstr\"om }
\def\des{de Sitter }

\def\section#1{\bigskip\noindent{\bf#1}\smallskip}

\def\PL#1{Phys.\ Lett.\ {\bf#1}}

\def\PR#1{Phys.\ Rev.\ {\bf#1}}\def\CQG#1{Class.\ Quantum Grav.\ {\bf#1}}
\def\NP#1{Nucl.\ Phys.\ {\bf#1}}

\def\ref#1{\medskip\everypar={\hangindent 2\parindent}#1}
\def\beginref{\begingroup
\bigskip
\centerline{\bf References}
\nobreak\noindent}
\def\endref{\par\endgroup}

\def\ef{e^{-2\f}}\def\emf{e^{-2\sqrt3\f_0}}
\def\ades{(anti)-de Sitter}


\magnification=1200
\baselineskip=18pt

{\nopagenumbers
\line{\hfil July 2009}
\vskip60pt
\centerline{\bf Exact solutions of dilaton gravity with \ades\ asymptotics}
\vskip60pt
\centerline{
{\bf S. Mignemi}\footnote{$^\ddagger$}{e-mail:
smignemi@unica.it}}
\vskip10pt
\centerline {Dipartimento di Matematica, Universit\`a di Cagliari}
\centerline{viale Merello 92, 09123 Cagliari, Italy}
\centerline{and INFN, Sezione di Cagliari}
\vskip80pt
\centerline{\bf Abstract}

\vskip10pt
{\noindent
We present a technique for obtaining \ssy\as\ades\ \bh solutions of dilaton
gravity with generic coupling to Maxwell field, starting from exact \af
solutions and adding a suitable dilaton potential to the action.
}
\vskip120pt\
}

\section{1. Introduction}
Charged dilatonic black holes have been largely studied in the context of the
low-energy limit of string theories [1]. The main characteristic of the
dilatonic models is the presence in the effective action of an exponential
coupling between the dilaton and the Maxwell field.
It results that this coupling gives rise to nontrivial scalar hair for black
holes, with a scalar charge independent from the other parameters of the
solution.

Dilatonic black holes have also been investigated in the case of more general
dilaton-Maxwell couplings [2], or when a cosmological constant or a scalar
potential of Liouville type are added to the action [3].
Also the possibility of multiple exponentially coupled scalar fields has been
considered [4].

On the other hand, there has been a renewed interest in the study of black-hole
solutions in theories with a cosmological constant, especially in the context of
the AdS/CFT correspondence [5].
In particular, the study of \ads black holes can give new insight in the
nonperturbative structure of some conformal models.
Moreover, it has been shown that, even in the case of minimally coupled scalar
fields, black holes with \ads asymptotics admit scalar hair, for a class
of scalar potentials [6].
This opens the possibility of phase transitions and breaking of gauge symmetry
near black hole horizons [7].

However, finding exact \bh solutions to dilatonic models with nonvanishing
cosmological constant is not trivial. In ref.\ [3] it was shown that \as\ads
solutions exist in the case of negative $\l$, but their analytical form is not
known. Moreover, in the $\l>0$ case, or when an exponential dilatonic potential
is added to the action, no \as\ades\ \bh solutions at all exist.
Nevertheless, the situation can change if more general dilatonic potential are
considered.

In fact, recently Gao and Zhang [8] have proposed a method that, starting from
an exact \ssy \af
solution of the Einstein-Maxwell-dilaton system, permits to obtain exact
asymptotically \des or \ads solutions of dilaton gravity exponentially coupled to
the Maxwell field, by the addition to the action of a suitable scalar potential.
Unfortunately, their derivation is rather involved
and it is not clear why it can give rise to consistent results.

In this letter we give a much simpler derivation of that result, which clarifies
this point, and generalize the method to the case of a generic coupling to the
Maxwell field. We then give some examples of application of the formalism to the
solutions found by Monni and Cadoni (MC) [2] in the \af case for a hyperbolic
dilaton-Maxwell coupling.

\section{2. Dilatonic \ades\ black holes.}
The proposal of [1] can be restated and generalized to arbitrary dilaton-Maxwell
coupling $W(\f)$, as follows: consider the action
$$I=\int\sqrt{-g}\ d^4x[R-2(\na\f)^2-W(\f)F^2],\eqno(1)$$
with \fe
$$R_\mn=2\na_\m\f\na_\n\f+2W(\f)\left(F^\r_{\ \m} F_{\r\n}-{1\over4}F^2
g_\mn\right)=0,$$
$$\na^2\f={1\over4}\,{dW\over d\f}\,F^2,\qquad \na_\m[W(\f)F^\mn]=0,\eqno(2)$$
and look for \ssy solutions of the form
$$ds^2=-U(r)\,dt^2+U^\mo(r)\,dr^2+R^2(r)\,d\O^2,\qquad\f=\f(r),\eqno(3)$$
with magnetic field
$$F=Q\sin\h\ d\h\wedge d\f.\eqno(4)$$

The \fe can then be written as
$${R''\over R}=-\f'^2,\eqno(5)$$
$${1\over R^2}(URR')'={1\over R^2}-W(\f)\,{Q^2\over R^4},\eqno(6)$$
$${1\over R^2}(R^2U\f')'=\ha\,{dW(\f)\over d\f}\ {Q^2\over R^4},\eqno(7)$$
and admit \af solutions. In particular, in the case $W(\f)=\ef$, the exact
solutions are well known, and read [1]
$$U(r)=1-{2M\over r},\qquad R^2(r)=r(r-2D),\qquad e^{-2\f}=e^{-2\f_0}
\left(1-{2D\over r}.\right)\eqno(8)$$
The solutions depend on three independent parameter. To simplify the following
discussion, we choose as independent parameters the mass $M$, the dilatonic
charge $D$ and the asymptotic value $\f_0$ of the dilaton. Of course, from a
physical point of view, the magnetic charge $Q$ is more relevant and could be
chosen instead of $\f_0$.
The magnetic charge is related to the other parameters by $Q^2=2MDe^{-2\f_0}$.

The solution displays a curvature singularity at $r=2D$ and a horizon at $r=2M$
and hence describes black holes if $M>D>0$. An independent scalar hair is
present, provided the magnetic charge is nonvanishing.

Define now a new metric given by
$$ds^2=-\left[U(r)-{\l\over3}R^2(r)\right]dt^2+
\left[U(r)-{\l\over3}R^2(r)\right]^\mo dr^2+R^2(r)d\O^2,\eqno(9)$$
where $U(r)$ and $R(r)$ are given by (8).
Under the conditions $M>D>0$, this is an \as(anti)-\des\bh with \cco $\l$.
More precisely, for $\l<0$, one has an \as \ads \bh with a single horizon, while for
$\l>0$, provided $\l D^2<3/4$, it results an \as \des \bh with one cosmological
horizon and one event horizon.

In this case, Gao and Zhang found that it is possible to add a scalar potential
$$V(\f)={2\over3}\l(2+\cosh[2(\f-\f_0)]),\eqno(10)$$
to the action (1), so that (9) is a solution of the ensuing field equations.

A drawback of this formalism is that $V(\f)$ depends explicitly on the asymptotic
value $\f_0$ of the scalar field, and hence no longer is an independent
parameter of the solution. It follows that for a given potential only two among
the charges $M$, $D$ and $Q$ (or $\f_0$) are independent, and in particular the
scalar charge is a function of $M$ and $Q$.
\bigskip
More generally, one can define the action
$$I=\int\sqrt{-g}\ d^4x[R-2(\na\f)^2-V(\f)-W(\f)F^2].\eqno(11)$$
With the ansatz (3)-(4), the \fe take the form
$${R''\over R}=-\f'^2,\eqno(12)$$
$${1\over R^2}(URR')'={1\over R^2}-W(\f)\,{Q^2\over R^4}-V(\f).\eqno(13)$$
$${1\over R^2}(R^2U\f')'=\ha\,{dW(\f)\over d\f}\ {Q^2\over R^4}
+{1\over 4}{dV(\f)\over d\f},\eqno(14)$$

Defining
$$U(r)=\bar U(r)-{\l\over3}R^2(r),\eqno(15)$$
where $\bar U(r)$ is a solution of (5)-(7), and exploiting the linearity of (13),
(14) in $U$, it is easy to see that the \fe (13), (14) reduce to
$${\l\over3}\,{1\over R^2}(R^3R')'=\ha\ V(\f),\eqno(16)$$
$${\l\over3}\,{1\over R^2}(R^4\f')'=-{1\over4}\ {dV(\f)\over d\f},\eqno(17)$$
while the field equation (12) is unchanged. Therefore, the solution for the
scalar field is identical to the \af one.
Then, inverting $\f(r)$, one can substitute in (16) or (17) and obtain $V$ as a
function of $\f$. Of course, one must show that the two equation are compatible.
In fact, substituting (12) in (16) and taking the derivative,
$${\l\over3}(12RR''+4RR'\f'^2+4R^2\f'\f'')=-{dV\over d\f}\f',\eqno(18)$$
and substituting again (12), one obtains (17).

\section{3. The generalization of MC solutions.}
As an example of application of this formalism, we consider the exact solutions
obtained by Monni and Cadoni [2] for the coupling $W(\f)=\cosh(2\a\f)$, when $\a=1$
or $\a=\sqrt3$. This coupling is invariant under the S-duality symmetry [9]
$\f\to-\f$, and therefore may be of interest for string theory.

In the case $\a=1$, the solution reads [2]
$$\bar U(r)={[r-(M-D)]^2-q^2\over r(r+2D)},\qquad R^2=r(r+2D),\qquad
e^{-2\f}=e^{-2\f_0}\,\left(1+{2D\over r}\right),\eqno(19)$$
where
$$q^2=M^2+D^2-2MD\coth2\f_0.\eqno(20)$$
The solution depends on three independent parameters: again we choose $M$, $D$ and
$\f_0$, in terms of which the magnetic charge reads
$$Q^2={2MD\over\sinh2\f_0}.\eqno(21)$$
It describes black holes if
$$M>D>0,\qquad\f_0>0,\qquad M^2+D^2-2MD\cosh2\f_0\ge0.\eqno(22)$$
(Solutions with $D<0$, $\f_0<0$ are related by duality to the previous ones).
In this case  a curvature singularity is located at the origin and is shielded by two
horizons located at $r_\pm=M-D\pm q$.

A metric function of the form (15) reads explicitly
$$U={-{\l\over3}\,r^4-{4\l\over3}\,Dr^3+(1-{4\l\over3}\,D^2)r^2-2(M-D)r-2MD(1-\coth2\f_0)
\over r(r+2D)}\ ,\eqno(23)$$
and applying the algorithm described above one can check that it is a solution of the
\fe for a potential
$$V(\f)={2\over3}\l(2+\cosh[2(\f-\f_0)]).\eqno(24)$$
Curiously, the potential has the same form as in the GHS case (10). Unfortunately, it
breaks the duality invariance $\f\to-\f$.

The solution is singular at $r=0$. If $\l<0$, it is \as \ads and, under the
conditions (22) it always exhibits two horizons. If $\l>0$, it is \as \des and for
$\l D^2<3/4$ it displays three horizons, otherwise a naked singularity emerges.
To the previous conditions, one must add the
request of positivity of the discriminant of the quartic polynomial in the numerator
of (23), in order for its roots to be real.
The expression of this condition is awkward and will not be reported here, but its
effect is to impose a limitation on the range of allowed values of the mass in function
of the other parameters.
The properties of the solution are analogous to those of the \ades-\RN black holes,
which are recovered in the limit $D=\f_0=0$, with $Q\ne0$.

\bigskip
In the case $\a=\sqrt3$, the MC solution reads [2]
$$\bar U(r)={(r-M)^2-q_0^2\over\sqrt{P_1P_2}},\qquad R^2(r)=\sqrt{P_1P_2},$$
$$e^{-2\sqrt3\f}=\emf\left({P_1\over P_2}\right)^{3/2},\eqno(25)$$
where
$$q_0^2={(D+M)^3F^2+(D-M)^3\over(D+M)F^2+(D-M)}\ge0,\eqno(26)$$
and
$$P_1=(r-D)^2-q_1^2,\qquad P_2=(r+D)^2-q_2^2.\eqno(27)$$
with
$$q_1^2={4D^2(D+M)F^2\over(D+M)F^2+(D-M)},\qquad
q_2^2={4D^2(D-M)\over(D+M)F^2+(D-M)}.\eqno(28)$$
Also this solution depends on three independent parameters, which we have chosen
to be the mass $M$, the scalar charge $D\over\sqrt3$ and $F=e^{2\sqrt3\f_0}$, with
$\f_0$ the asymptotic value of the scalar field. The magnetic charge $Q$ is
then related to the other charges by the constraint
$$Q^2={4D(D^2-M^2)F\over(D+M)F^2+(D-M)}.\eqno(29)$$
The requirements $q_0^2\ge0$, $Q^2>0$ yield
$$F^2\le\left({M-D\over M+D}\right)^3.\eqno(30)$$
Under these conditions, $q_1^2<0$, $q_2^2>0$, and hence $P_1$ has no real roots,
while a curvature singularity occurs at the greatest root of $P_2$, namely
$r_0=-D+q_2$.
This is shielded by two horizons placed at $r_\pm=M\pm q_0$ iff $M>D>0$, $F<1$
(again, solutions with $D<0$, $F>1$ are obtained by duality).

Let us now consider a metric of the form (15). The metric function $U(r)$ reads
explicitly
$$U=\,{-{\l\over3}r^4+(1+2\l D^2)r^2-2[M+{\l\over3}D(q_1^2-q_2^2)]r
+M^2-q_0^2-{\l\over3}(3D^4-q_1^2q_2^2)\over R^2(r)}.\eqno(31)$$
Applying the formalism developed above, it is easy to check that $U(r)$
solves the \fe for a potential
$$V=2\l\cosh{2\over\sqrt3}(\f-\f_0).\eqno(32)$$
Also in this case the duality invariance is broken.

A curvature singularity still occurs at $r_0$, and
the solution describes black holes under the same restriction of (30).
Moreover, in the \ads case, $\l<0$, one must require
$$|\l|\ge{3\over4}\,{M+D-q_2\over q_2^2(q_2-2D)},\eqno(33)$$
obtaining a \bh with two horizons.

Analogously, in the \des case, a \bh with three horizon exists if
$$\l\le{1\over2q_2(q_2-2D)}.\eqno(34)$$

As in the previous case, one must also impose a condition that ensures the existence
of real roots of the polynomial in the numerator of (30), and gives rise to
restrictions on the allowed values of the mass.
All other values of the parameters lead to naked singularities.
Also in this case, the structure of the solution is analogous to that of the
\ades-\RN black holes. These are recovered when $D=0$, $F=1$.

The difference between the properties of the \as\ades\ GHS solution compared to the
RN or MC ones is due to the degree of the polynomial appearing in the metric function.
It appears that the GHS is a degenerate limit case, in which one of the roots of $\bar U$
coincides with one root of $R^2$, while the generic case is of \RN type.

Electric charged \bhs for the dilaton-Maxwell coupling $W(\f)=1/\cosh2\a\f$ can be
obtained as usual by duality from the magnetic ones. The metric maintains its form,
while $\f\to-\f$.
The CM solutions are already invariant under this transformation, and hence also the
scalar potential is unchanged, except for the sign of $\f_0$.

Finally we notice that, as in the \RN case, the MC solutions can present degenerate
horizons when two of the horizons merge. This happens when the discriminant discussed
above vanishes.

\section{4. Conclusions.}
We have generalized a method proposed in [8] to obtain \ssy\as\ades\ \bh solutions
in dilatonic models nonminimally coupled to the Maxwell field, by the addition of a
suitable scalar potential.
In particular, we have found that in the case of hyperbolic coupling, the dilatonic
potential is also hyperbolic.
In this case, the metric has the same causal structure as that of the \ades-\RN
solution of general relativity.

It is interesting to notice that, while in the \af case the asymptotic value of the
dilaton is free, in the \ades\ case the presence of a dilaton potential fixes its value,
and in general the solutions only depend on two parameters. In this case, the dilatonic
charge is therefore determined by the mass and the Maxwell charge.
This is in accordance with the fact that, according to well known no hair theorems for
minimally coupled \ads black holes, the asymptotic value of the scalar field must coincide
with an extremum of the potential [10].
It would be interesting to investigate if this fact could lead to no-hair theorems
for nonminimally coupled models. In particular, it is not clear if the solutions obtained
are unique or if the absence of an independent scalar charge is an artifact of the way in
which the solutions have been constructed.
\bigbreak

\section{Acknowledgements}
\noindent{I wish to thank M. Cadoni and P. Pani for useful comments.}

\beginref
\ref [1] G.W. Gibbons and K. Maeda, \NP{B298}, 741 (1988);
D. Garfinkle, G.T. Horowitz and A. Strominger, \PR{D43}, 3140 (1991);
C.F.E. Holzhey and F. Wilczek, \NP{B380}, 447 (1992).
\ref [2] S. Monni and M. Cadoni, \NP{B466}, 101 (1996).
\ref [3] S.J. Poletti and D.L. Wiltshire, \PR{D50}, 7260 (1994);
S.J. Poletti, J. Twamley and D.L. Wiltshire, \PR{D51}, 5720 (1995).
\ref [4] M. Cadoni and S. Mignemi, \PR{D48}, 5536 (1993);
S. Mignemi, \PR{D62}, 024014 (2000);
S. Mignemi  and D.L. Wiltshire, \PR{D70}, 124012 (2004).
\ref [5] J.M. Maldacena, Adv. Theor. Math. Phys. {\bf2}, 231 (1998);
E. Witten, Adv. Theor. Math. Phys. {\bf2}, 253 (1998).
\ref [6] T. Torii, K. Maeda and M. Narita, \PR{D64}, 040407 (2001).
\ref [7] S.S. Gubser, \CQG{22}, 5121 (2005); \PR{D78}, 065034 (2008).
\ref [8] C.J. Gao and S.N. Zhang, \PR{D70}, 124019 (2004).
\ref [9] A. Font, L.E. Iba\~nez, D. L\"ust and F. Quevedo, \PL{B249}, 35 (1990).
\ref [10] T. Torii, K. Maeda and M. Narita, \PR{D64}, 044007 (2001);
T. Hertog, \PR{D74}, 084008 (2006).

\endref
\end
J.D. Bekenstein, \PR{D 5}, 1239 (1972); \PR{D 5}, 2403 (1972);
S.J. Poletti, J. Twamley and D.L. Wiltshire,
\PR{D51}, 5720 (1995); 
\CQG{12}, 1753 (1995), (E) {\bf12}, 2355 (1995); 
D.L. Wiltshire,
J.\ Austral.\ Math.\ Soc.\ {\bf B41}, 198 (1999). 
S. Mignemi and D.L. Wiltshire, \PR{D46}, 1475 (1992);